\documentstyle[prd,aps,epsfig,amssymb,amsfonts,tighten]{revtex}

\begin{document} \draft \def\si{\sigma}
\twocolumn[\hsize\textwidth\columnwidth\hsize\csname @twocolumnfalse\endcsname
\title{\hfill OKHEP--01--10\\
Remark on the perturbative component of inclusive $\tau$-decay}

\author{K.A. Milton}
\address{Department of Physics and Astronomy, University of Oklahoma, Norman, OK 73019 USA}
\author{I.L. Solovtsov and O.P. Solovtsova}
\address{ Bogoliubov Laboratory of Theoretical Physics, JINR,
         Dubna, 141980 Russia}
\preprint{OKHEP--01--10}
\date{\today}
\maketitle

\begin{abstract}
In the context of the inclusive $\tau$-decay, we analyze various forms of
perturbative expansions which have appeared as modifications of the original
perturbative series. We argue that analytic perturbation theory, which
combines renormalization-group invariance and $Q^2$-analyticity, has
significant merits favoring its use to describe the perturbative component
of $\tau$-decay.
\end{abstract}
\vskip2pc]

A perturbative  approximation in quantum chromodynamics as a rule cannot be
exhaustive in the low energy region of a few GeV and a nonperturbative
component has to be included. The reliability of extracting
nonperturbative parameters from data is connected with uncertainties in
the perturbative description of a process arising from
the inevitable truncation of
the perturbation theory (PT) series. The initial perturbative series that is
obtained after the renormalization procedure is not the final product of
the theory. This series can be modified and its properties can be improved
on the basis of additional information coming from general properties of the
quantity under consideration. In this note we consider various descriptions of
the perturbative component in the context of an analysis of the
inclusive  decay of the $\tau$ lepton.
We will discuss merits and drawbacks of the series
expansion for the $R_\tau$-ratio in terms of powers of the
parameter $\alpha_s(M_\tau^2)$
\cite{Braaten88}, the prescription of Ref.~\cite{LP92} which uses a contour
representation, and the approximation based on the analytic approach
proposed in Ref.~\cite{SS96-97}.

The main object in a description of the hadronic decay of the $\tau$-lepton
and of many other physical processes is the correlator $\Pi(q^2)$ or the
corresponding Adler function $D(-q^2)=-q^2{d\Pi(q^2)/}{dq^2}$. The analytic
properties of the $D$-function are contained within the relation
\begin{equation}\label{D-funct2}
D(Q^2)=Q^2\int_0^{\infty}\frac{ds}{{(s+Q^2)}^2}R(s)\,,
\end{equation}
where $R(s)={\rm{Im}}\Pi(s)/\pi$. According to this equation, the
$D$-function is an analytic function in the complex $Q^2$-plane with a cut
along the negative real axis.

After renormalization, the perturbative expansion of the $D$-function has
the form of a power series in the expansion parameter
$a_\mu=\alpha_s(\mu^2)/\pi$. In the massless case the series has the form
\begin{eqnarray}\label{D-PT-init}
&&D(Q^2)\propto1+a_\mu\,d_{1,0}+a^2_\mu\left(d_{2,0}+d_{2,1}\ln\frac{Q^2}{\mu^2}\right)
\nonumber  \\
&&+a^3_\mu\left(d_{3,0}+d_{3,1}\ln\frac{Q^2}{\mu^2}+d_{3,2}\ln^2\frac{Q^2}{\mu^2}\right)+\cdots.
\end{eqnarray}
As is well known, this expression is unsatisfactory both from
theoretical and practical viewpoints. Any partial sum of it is not
renormalization-group invariant and logarithms in the coefficients lead to
an ill-defined behavior in both infrared and ultraviolet regions.

The modification of the initial representation (\ref{D-PT-init}) based on
renormalization-group invariance reads
\begin{eqnarray}\label{D-PT-RG}
D(Q^2)&\propto&1+\bar{a}(Q^2)d_{1,0}+\bar{a}^2(Q^2)d_{2,0}  \nonumber \\
&+&\bar{a}^3(Q^2)d_{3,0}+\cdots\,,
\end{eqnarray}
where $\bar{a}(Q^2)$ is the running coupling. This commonly used
modification removes some of the undesirable features of the expansion
(\ref{D-PT-init}). A partial sum of the series (\ref{D-PT-RG}) is now
$\mu$-independent. The log-terms in the coefficients of
Eq,~(\ref{D-PT-init}) have been summed into the running coupling and the
series (\ref{D-PT-RG}) can now be used in the ultraviolet region. However,
the correct analytic properties of the partial sum of Eq.~(\ref{D-PT-init}),
the principal merit of this expansion, are no longer valid due to unphysical
singularities of the perturbative running coupling.

An analytic approach proposed in Ref.~\cite{SS96-97} gives a possible
resolution of this problem. The series (\ref{D-PT-RG}) has been constructed
from that in Eq.~(\ref{D-PT-init}) by using additional information of a
general type---the renormalization-group invariance of the quantity under
consideration. The analytic approach makes the next logical step in the
modification of the perturbative expansion by bringing into consideration
additional general principles of the theory which are reflected in the
$Q^2$-analyticity. In the framework of analytic perturbation
theory (APT), the series has the form of a nonpower expansion \cite{MSS97}
\begin{eqnarray} \label{D-APT}
D(Q^2)&\propto&1+{\cal{A}}_1(Q^2)\,d_{1,0}+{\cal{A}}_2(Q^2)\,d_{2,0}  \nonumber \\
&+&{\cal{A}}_3(Q^2)\,d_{3,0}+\cdots\,,
\end{eqnarray}
where ${\cal{A}}_k(Q^2)$ are analytic functions in the complex $Q^2$-plane
with a cut along the negative real axis. The expansion (\ref{D-APT})
maintains not only correct analytic properties of the partial sum, but leads
to some new remarkable features. For example, within the APT the
renormalization scheme (RS) dependence of results obtained, caused by the
inevitable truncation of the series, is reduced drastically (see details and
applications to various processes in
Refs.~\cite{SS98,Sh99,SS99tmp,MSS98Bj,MSS99GLS,MSSY00}). Moreover, the
analytic approach allows one to give a self-consistent definition of
the perturbative expansion in the timelike region~\cite{MS97}. The APT
representation of the $R$-ratio in the process of $e^+e^-$ annihilation into
hadrons, defined in the timelike region, like the $D$-function, defined in
the spacelike region, has the form of a nonpower expansion
\begin{eqnarray} \label{R-APT}
R(s)&\propto& 1+{\mathfrak{A}}_1(s)\,d_{1,0}+{\mathfrak{A}}_2(s)\,d_{2,0}  \nonumber \\
&+&{\mathfrak{A}}_3(s)\,d_{3,0}+\cdots\, .
\end{eqnarray}
Both expansions, Eqs.~(\ref{D-APT}) and (\ref{R-APT}), are related term by term
by
Eq.~(\ref{D-funct2}). The functions ${\cal{A}}_k$ and ${\mathfrak{A}}_k$ are
close  to each other, although they do not
coincide \cite{MS97,MOlSol98,Shirkov01a,Shirkov01b}.

The functions $D(Q^2)$ and $R(s)$ can be written in terms of an effective
spectral function $\rho(\sigma)$ \cite{MS97}:
\begin{equation}\label{d-rho}
D(Q^2)\propto1+\frac{1}{\pi}\int_0^{\infty}\,d\sigma \frac{\rho(\sigma)}{\sigma+Q^2}\,,
\end{equation}
\begin{equation}\label{r-rho}
R(s)\propto1+\frac{1}{\pi}\,\int_s^{\infty}\frac{d\sigma}{\sigma}\,
\rho(\sigma)\,.
\end{equation}
The function $\rho(\sigma)$ is calculated by taking the discontinuity of
$D(Q^2)$ across the cut.

The leading-order expansion functions in the spacelike and timelike regions are
\begin{equation}         \label{a1E}
{\cal{A}}_1(Q^2)\,=\,\frac{1}{\beta_0}
\left[\frac{1}{\ln(Q^2/\Lambda^2)} + \frac{\Lambda^2}{\Lambda^2-Q^2} \right]\, ,
\end{equation}
\begin{equation}         \label{a1M}
{\mathfrak{A}}_1(s)\,=\,\frac{1}{\beta_0}\left[
\frac{1}{2}-\frac{1}{\pi}\arctan\frac{\ln( s/\Lambda^2)}{\pi} \right] \,.
\end{equation}
The expression (\ref{a1E}) contains the standard logarithmic term which
coincides with the perturbative expression having a ghost pole at
$Q^2=\Lambda^2$. The second term in Eq.~(\ref{a1E}) (which appears automatically
from the K\"all\'en--Lehmann representation) has a non-logarithmic power
structure, cancels the ghost pole, and provides expression (\ref{a1E}) with
the correct analytic properties.
This term has a short distance origin unrelated to the operator product expansion
and, therefore, is not inconsistent with it (see, for instance, the discussion
in Ref.~\cite{Grunberg97}).
Rewritten in terms of the PT
running coupling the second term in Eq.~(\ref{a1E}) has the structure
of the type $\exp(-1/\bar{a})$ and it
therefore makes no contribution to the perturbative series. Thus, the true
analytic properties of the function (\ref{a1E}) are restored by terms which
are invisible in the perturbation expansion, which terms, nevertheless, are
felt by the dispersion relation, even for the spectral function calculated
perturbatively. The regularity of the function (\ref{a1M}), defined in the
timelike region, has another origin. Its perturbative expansion contains
only logarithmic terms proportional to powers of $\pi^2/\ln^2(s/\Lambda^2)$.
The $\pi^2$-terms are usually incorporated into the coefficients in the
perturbative expansion of $R(s)$ without changing the expansion parameter,
i.e., the form of the running coupling in the timelike region is taken to be
the same as in the spacelike region. In any order of that expansion the
relation (\ref{D-funct2}) will be violated.

The experimentally measured $R_\tau$-ratio of hadronic to leptonic widths is given by
\begin{equation} \label{R_tau-init}
R_{\tau}=2 \int^{M_\tau^2}_0 {ds \over M_\tau^2 } \,
\left(1-{s \over M_\tau^2}\right)^2 \left(1 + 2 {s \over M_\tau^2}\right)R(s)\,.
\end{equation}
This expression can be represented in the form of a contour integral in
terms of the Adler $D$-function \cite{Braaten88}
\begin{equation} \label{R_tau-contour}
R_{\tau}\, =\, \frac{1}{2\pi {\rm i}} \, \oint_{|z|=M_\tau^2} \frac{d\,z}{z}\,
\left(1-{z \over M_\tau^2}\right)^3 \left(1 + {z \over M_\tau^2}\right) D(-z) \, .
\end{equation}
The integration in Eq.~(\ref{R_tau-init}) contains an interval over timelike
momentum which extends down to small $s$ and, therefore, cannot directly be
calculated if one applies the standard parameterizations of $R(s)$ in terms
of  the singular PT running coupling. At first glance the contour
representation (\ref{R_tau-contour}) solves this problem because the
integration over the circle is well-defined even if the $D$-function is
written in terms of $\alpha_s(Q^2)$. However, in order to perform this
transformation self-consistently, it is necessary to maintain required
analytic properties of the $D$-function, which are violated in the framework
of standard PT with a singular running coupling.

Two approaches are usually used to describe the perturbative component in
inclusive $\tau$-decay. In the first one~\cite{Braaten88}, called the
fixed-order perturbation theory (FOPT), the perturbative QCD contribution
$\delta_\tau$ to $R_\tau$ is represented in the form of a power expansion in
the parameter $\bar{a}\left(M_\tau^2\right)$. The three-loop approximation
reads
\begin{equation} \label{delta_I}
\delta_{\tau}^{\rm FOPT} =\bar{a}\left(M_\tau^2\right)\, +
K_{1}\,\bar{a}^2\left(M_\tau^2\right) + K_{2}\,\bar{a}^3\left(M_\tau^2\right)\, ,
\end{equation}
where in the $\overline{\rm{MS}}$-scheme for three active flavors the
coefficients are $K_1=5.2023$ and $K_2=26.366$.

Another PT form is obtained if the perturbative expansion (\ref{D-PT-RG}) is
substituted into the contour integral (\ref{R_tau-contour}). This gives the
so-called contour-improved perturbation theory (CIPT)
representation~\cite{LP92}
\begin{equation}  \label{delta_II}
\delta_{\tau}^{\rm CIPT} = A^{(1)}\left(M_\tau^2\right) + d_1
\,A^{(2)}\left(M_\tau^2\right)+ d_2\,A^{(3)}\left(M_\tau^2\right)
\end{equation}
with
\begin{eqnarray} \label{A-n}
A^{(n)}\left(M_\tau^2\right)&=&\frac{1}{2\pi{i}}
\oint_{|z|=M_{\tau}^2}\frac{dz}{z}\\ &\times&
{\left(1-\frac{z}{M_{\tau}^2}\right)}^3 \left(1+\frac{z}{M_{\tau}^2}\right)
\bar{a}^n(-z)\,, \nonumber
\end{eqnarray}
where $d_1=1.640$ and $d_2=6.371$ in the $\overline{\rm{MS}}$-scheme.

Both expressions, Eqs.~(\ref{delta_I}) and (\ref{delta_II}), are widely used in
the analysis of $\tau$-decay data. However, their status is different. The
formula (\ref{delta_I}) can be obtained self-consistently. In
Eq.~(\ref{R_tau-init}) one has to use for $R(s)$ the initial perturbative
approximation, like Eq.~(\ref{D-PT-init}) for $D$-function, with the
expansion parameter $a_\mu$. Then, after integration over $s$, the
logarithmic terms containing $\ln(M_\tau^2/\mu^2)$ are removed by setting
$\mu^2=M_\tau^2$. The same result is obtained if the contour representation
(\ref{R_tau-contour}) is used and the $D$-function is taken in the form
(\ref{D-PT-init}) which preserves the required analytic properties. As for
the representation (\ref{delta_II}), it will be consistent
 with Eqs.~(\ref{R_tau-init})
and (\ref{R_tau-contour}) if $\bar{a}(z)$ has analytic properties of the
K\"all\'en--Lehmann type. The use of the standard
running coupling with unphysical singularities in Eq.~(\ref{A-n}) breaks
this consistency.

In the framework of APT, where the $Q^2$-analyticity is maintained, the
expression for $\delta_\tau$ can be obtained both from the initial formula
(\ref{R_tau-init}) and from the contour representation
(\ref{R_tau-contour}). In terms of the effective spectral function it has
the form~\cite{MSS97}
\begin{eqnarray} \label{delta_an}
\delta_{\tau}^{\rm{APT}}&=&\frac{1}{\pi}\int_0^{\infty}\frac{d\si}{\si}\rho(\si)
\\
&-&\frac{1}{\pi}\int^{M_{\tau}^2}_0\frac{d\si}{\si}{\left(1-\frac{\si}{M_{\tau}^2}\right)}^3
\left(1+\frac{\si}{M_{\tau}^2}\right)\rho (\si)\,. \nonumber
\end{eqnarray}

           \begin{figure}[hpt]
\centerline{\epsfig{file=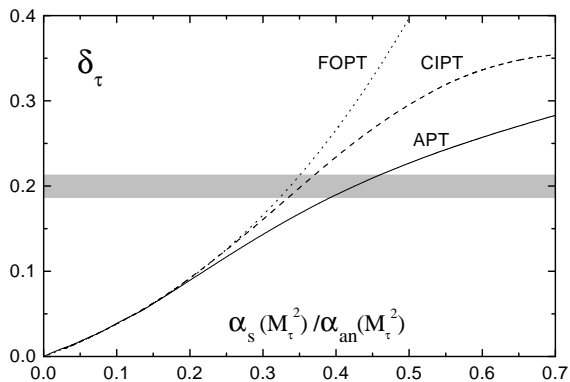,width=9.1cm}} %
\caption{\sl{Behavior of the QCD perturbative contribution to $R_\tau$ as a
function of
$\alpha_s(M_\tau^2)$ (for FOPT and CIPT) or of
$\alpha_{\rm{an}}(M_\tau^2)$ (for APT) in the
$\overline{\rm{MS}}$ renormalization scheme.}}
         \label{delta-0}
         \end{figure}

In Fig.~\ref{delta-0}, we compare $\delta_{\tau}$ as a function of the
three-loop running coupling in the cases of FOPT, CIPT, and APT.
The running coupling in the Euclidean region is defined by
\begin{equation}
\alpha(Q^2)=\frac1\pi\int_0^\infty\frac{d\sigma}{\sigma+Q^2}\mbox{Im}\left[
\alpha_{\rm pt}(-\sigma-i\epsilon)\right],
\end{equation}
which in lowest order coincides with Eq.~(\ref{a1E}).
This plot
corresponds to calculations in the $\overline{\rm{MS}}$-scheme. The
difference between these functions is negligible at sufficiently small
$\alpha_s(M_\tau^2)$ and becomes substantial with larger values of the
coupling. The shaded area reflects the value
$\delta_\tau=0.200\pm0.013$~\cite{Pich00} which corresponds to the
experimental measurement for the non-strange channel of the $\tau$-data
\cite{ALEPH98,OPAL99}.

A discussion of the perturbative contribution to inclusive $\tau$-decay has
been given in Ref.~\cite{G-Ioffe_Z}, in which it has been claimed that in
order to describe the experimental data for $\tau$-decay the value of
$\alpha_{\rm{an}}(M_\tau^2)$ should be taken in the range of $1.5$--$2.0$.
This conclusion contradicts our result. Fig.~\ref{delta-0} clearly
demonstrates that in order to reproduce the experimental data such large
values of $\alpha_{\rm{an}}(M_\tau^2)$ are not required. Moreover, in the
APT approach, the value of the analytic running coupling
$\alpha_{\rm{an}}(Q^2)$ is bounded from above and cannot exceed the infrared
limiting value $\alpha_{\rm{an}}\leq\pi/\beta_0\simeq1.4$ \cite{SS96-97}.
Beyond this, in Ref.~\cite{G-Ioffe_Z}, it was noted that this impossibly
large value $\alpha_{\rm{an}}(M_\tau^2)$ corresponds to
$\alpha_{s}(M_{Z}^2)\simeq0.15$. However, in order to obtain the value of
$\alpha_s$ at the $Z$-boson mass scale, the region of five active quarks
should be approached by applying a special procedure of matching from the
three-quark region \cite{MOlSol98,Shirkov01a}. Corresponding estimates have
been given in Ref.~\cite{MSSY00}. The large value of the scale parameter
$\Lambda$ extracted in a pure APT analysis of the $\tau$-data indicates that
nonperturbative effects are not negligible. As it has been demonstrated in
Ref.~\cite{MSS01}, the light $D$-function corresponding to the non-strange
vector channel $\tau$-data can be described by using
reasonable effective quark masses
with $\Lambda_{\overline{\rm{MS}}}\simeq420$~MeV.

As has been emphasized in Ref.~\cite{G-Ioffe_Z}, the merit of the contour
representation is that it produces expressions for quantities in the
physical region that are not expansions in the parameter
$\pi/\ln(\,s/\Lambda^2)$, not a small quantity in the intermediate energy
region. In particular, one can write the formula for $R(s)$ [see Eq.~(11)
from Ref.~\cite{G-Ioffe_Z}] which contains the expression on the right-hand
side of Eq.~(\ref{a1M}). Although we agree with the authors of
Ref.~\cite{G-Ioffe_Z} that it is important to sum up the $\pi^2$-terms in
the region of intermediate energies (see Ref.~\cite{Shirkov01b}) and that it
is preferable to use an expression which sums up the $\pi^2$-contributions
rather than the asymptotic expression $\propto1/\ln(s/\Lambda^2)$, we note
the following: The expression discussed in Ref.~\cite{G-Ioffe_Z} is simply
associated with the APT timelike function (\ref{a1M}) which unambiguously
leads to the Euclidean function (\ref{a1E}). So, we conclude that the
analytic approach provides a consistent form of expressions of the type
(\ref{a1M}) in the timelike region, whose evident advantage is the summation
of the $\pi^2$-contributions into a regular function, and
results in the corresponding analytic expressions for the $D$-function in the
spacelike region, where unphysical singularities are cancelled by functions
of a nonlogarithmic type vanishing in a perturbative expansion. In other
words, summation of the $\pi^2$-contributions in the s-channel produces
power (nonlogarithmic in $Q^2$) contributions in the t-channel that ensure
the analytic properties by canceling unphysical singularities in the
logarithmic terms.

\begin{figure}[hpt]
\centerline{\epsfig{file=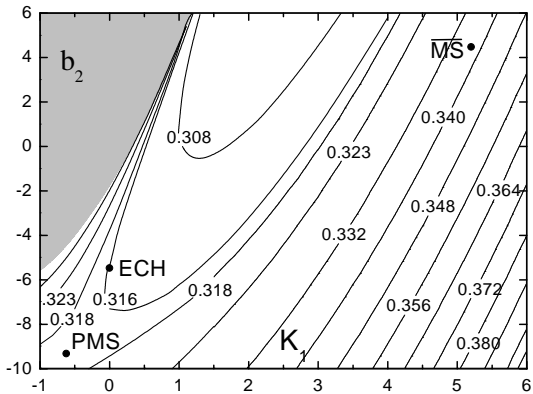,width=9.5cm}} %
\caption{\sl{Contour plot of $\alpha_{s}^{\overline{\rm{MS}}}(M_{\tau}^2)$
extracted from the $\tau$-data by using different renormalization schemes
in the FOPT approach.
        }}
         \label{alpha-contour}
         \end{figure}

A significant source of theoretical uncertainty of the results obtained
arises from the RS dependence due to the inevitable inclusion of only a
finite number of terms in the PT series. There are no general principles
that give preference to a particular renormalization scheme and the
stability of  the result has to be studied at least with respect to
a relevant class of schemes. A virtue of the APT approach is its higher
stability over a wide range of RS~\cite{MSSY00}.
(A similar observation has been made in the context of deep-inelastic
scattering sum rules \cite{MSS98Bj,MSS99GLS}.)
In contrast, results
obtained in the framework of FOPT have a strong RS dependence. Indeed, if
one performs calculations in some scheme and then by using a RS transformation
extracts the value of $\alpha_{s}^{\overline{\rm{MS}}}(M_{\tau}^2)$ in the
commonly used ${\overline{\rm{MS}}}$-scheme, we find that this value
possesses a strong uncertainty. This fact is shown in
Fig.~\ref{alpha-contour}. A point on the $(K_1,b_2)$-plane corresponds to a
RS, where $b_2=\beta_2/\beta_0$ is the ratio of the three-loop and one-loop
coefficients of the $\beta$-function. In the shaded area there are no
appropriate solutions of the cubic equation (\ref{delta_I}). We indicate the
popular ${\overline{\rm{MS}}}$ scheme and the optimal schemes: ECH, which
recently has been applied to the $\tau$-decay \cite{Pivovarov00}, based on
an effective charge method \cite{Grunberg-Dhar-Gupta}, and PMS based on a
principle of minimal sensitivity \cite{Stevenson-PMS}. A straightforward way
of extracting the value of the coupling based solely on the
${\overline{\rm{MS}}}$ scheme gives
$\alpha_s^{\overline{\rm{MS}}}(M_\tau^2)=0.337\pm0.008$. An analysis which
takes into account the RS ambiguity leads to a much larger error in
$\alpha_s(M_\tau^2)$.  The reason for the enhanced RS stability of the APT
scheme is its perturbative stability---the leading order term captures
the bulk of the QCD correction.

In summary, we have made a comparative analysis of the merits and drawbacks
of three forms of perturbative approximations in the context of the
application of QCD to the inclusive decay of the $\tau$-lepton. Two of them,
the FOPT and CIPT, use,  in one way or another, the running coupling with
unphysical singularities to parameterize $R_\tau$. However, the
justification of these schemes is rather different. The FOPT has a more
solid foundation than the CIPT, which is simply self-contradictory. However,
due to the strong RS dependence, the FOPT leads to a large uncertainty in
extracting QCD parameters from the data. We have put forward arguments in
favor of APT. Calculations in the framework of this approach are
self-consistent and considerably reduce the theoretical uncertainty of the
results obtained; this is important in extracting the nonperturbative
component of the QCD description of the process.

The authors would like to thanks D.V.~Shirkov for interest in this work and
useful discussions. Partial support of the work by the U.S. Department of
Energy, grant number DE-FG03-98ER41066, and by the RFBR, grants 99-01-00091,
99-02-17727, and 00-15-96691, is gratefully acknowledged.


\end{document}